\documentclass[prd,12pt]{revtex4}

\usepackage{epsfig}
\usepackage{graphicx}
\usepackage{dcolumn}
\usepackage{amsmath}
\usepackage{latexsym}
\usepackage{color}
\usepackage{subfig}

\usepackage{footnote}
\usepackage{hyperref}
\hypersetup{
    colorlinks=true,
    linkcolor=blue,
    citecolor=blue,
    filecolor=blue,      
    urlcolor=blue,
    hyperfootnotes=true
}
 
\begin{document}

\title{Generalised Jordan map, symplectic transformations and Dirac's representation of the $3+2$ de Sitter group}  

\author{Rabin Banerjee\footnote{
\href{mailto:rabin@bose.res.in}{rabin@bose.res.in}}}
\affiliation{Department of Theoretical Sciences, S. N. Bose National Centre for Basic Sciences,\\ Block-JD, Sector-III, Salt Lake City,\\ Kolkata 700106, India}

\begin{abstract}
\noindent In his 1963 paper, Dirac \cite{pamd} gave \textit{`A Remarkable Representation of the $3+2$ de-Sitter Group'}. We reproduce this representation using a generalised Jordan map which is motivated by the infinitesimal symplectic transformations related to the four dimensional symplectic 
group $Sp(4)$. A physical picture of Dirac's representation is also discussed.  

\end{abstract}

\maketitle


\section{Introduction}
\noindent Paul Dirac, in 1963, published a paper on \textit{A Remarkable Representation of the $3+2$ de-Sitter Group} \cite{pamd}. He further showed that, among the infinitesimal generators of this group, there are four cyclic ones while the rest are hyperbolic. The representation is very simple with the wavefunctions depending on two variables. \\ 
 \hspace*{2mm} The de-Sitter group is an important space-time symmetry in both particle physics and general relativity. It was initially introduced into physics for describing curved background \cite{hpr}, leading to significant cosmological consequences. It has several interesting subgroups and is also the starting point for building representations of the Poincare group for relativistic particles \cite{epw1}. The algebra presented in \cite{pamd} has found many applications \cite{kn}, especially in providing an explicit example of connecting the standard Schroedinger approach to the quantum mechanics with Wigner's phase space approach \cite{epw2}. Its mathematical construction has been used for building the two mode squeezed states in quantum optics \cite{hkn, hkny, ymk}. Thus, it has an important role in contemporary physics.\\
 \hspace*{2mm}  In this paper we give a simple derivation of Dirac's representation that is based on a generalisation of the well known Jordan-Schwinger map that yields a realisation of $SU(2)$ generators. This generalisation is motivated by the fact that the algebraic structure of Dirac's representation may be understood by infinitesimal symplectic transformations related to the group $Sp(4)$. It is possible to reproduce the complete structure of the representation \cite{pamd} using these transformations. Finally, some other features are discussed that provide new insights into this representation. \\ 
\hspace*{2mm} In section II, we briefly review the salient features of Dirac's remarkable representation. This is followed by showing, in section III, that this representation may be obtained from the $Sp(4)$ group using infinitesimal symplectic transformations. In section IV, we show that all results may be simply obtained by using a generalised Jordan-Schwinger type map. Section V discusses some physical aspects of Dirac's representation. Finally, Section VI contains our conclusion, including possible future directions.
\section{The Representation} 
\noindent Dirac considered a pair of coupled harmonic oscillators. The ladder operators, denoted by $a_{i}$ and $a_{i}^{\dagger}$ satisfy the basic commutation relations, $$[a_{i}, a_{j}] = [a_{i}^{\dagger}, a_{j}^{\dagger}] = 0$$
\begin{equation}
~~~~~~~~~~~~~~~~~~~~~~~~~~~~~~~~~~~~~~[a_{i}, a_{j}^{\dagger}] = \delta_{ij} ~~~~~~~~~~~~~~~~~~~~~~~~~~~~~~~~~ (i,j = 1,2).
\label{eq1}
\end{equation} 
\noindent Now sixteen quadratic operators can be constructed from these $a_{i}$ and $a_{i}^{\dagger}$. However, because of the relations (\ref{eq1}), only ten of these are independent. Appropriate linear combinations of these were used to define the following ten operators \cite{pamd},  $$ L_{1} = \dfrac{1}{2}(a_{1}^{\dagger}a_{2} + a_{2}^{\dagger}a_{1}) ~~ , ~~ L_{2} = \dfrac{i}{2}(a_{2}^{\dagger}a_{1} - a_{1}^{\dagger}a_{2}) $$ $$L_{3} = \dfrac{1}{2}(a_{1}^{\dagger}a_{1} - a_{2}^{\dagger}a_{2}) ~~ , ~~ H = \dfrac{1}{2}(a_{1}^{\dagger}a_{1} + a_{2}a_{2}^{\dagger})  $$ $$  K_{1} = -\dfrac{1}{4}(a_{1}^{\dagger}a_{1}^{\dagger} + a_{1}a_{1} - a_{2}^{\dagger}a_{2}^{\dagger} - a_{2}a_{2}) ~~~~~~~~~~~$$  $$  K_{2} = \dfrac{i}{4}(a_{1}^{\dagger}a_{1}^{\dagger} +  a_{2}^{\dagger}a_{2}^{\dagger} - a_{1}a_{1} - a_{2}a_{2}) ~~~~~~~~~~~~~$$  $$ K_{3} = \dfrac{1}{2}(a_{1}^{\dagger}a_{2}^{\dagger} + a_{1}a_{2})~~~~~~~~~~~~~~~~~~~~~~~~~~~~~~~~$$
 $$  B_{1} = -\dfrac{i}{4}(a_{1}^{\dagger}a_{1}^{\dagger} - a_{1}a_{1} - a_{2}^{\dagger}a_{2}^{\dagger} + a_{2}a_{2}) ~~~~~~~~~~~$$ $$  B_{2} = -\dfrac{1}{4}(a_{1}^{\dagger}a_{1}^{\dagger} +  a_{2}^{\dagger}a_{2}^{\dagger} + a_{1}a_{1} + a_{2}a_{2}) ~~~~~~~~~~~$$ 
\begin{equation}
B_{3} = \dfrac{i}{2}(a_{1}^{\dagger}a_{2}^{\dagger} - a_{1}a_{2}) ~~~~~~~~~~~~~~~~~~~~~~~~~~~~~~~~
\label{eq2}
\end{equation}

\noindent The first four $(L_{1}, L_{2}, L_{3}, H)$ are cyclic rotations with $H$ separate from the other three, since it commutes with all of them. The remaining six correspond to hyperbolic rotations. These operators provide a representation of the $3+2$ de-Sitter group. This is the group of rotations of five real variables $q_{1}, q_{2}, q_{3}, q_{4}, q_{5}$ that keeps the quadratic form, $$q_{1}^{2}+q_{2}^{2}+q_{3}^{2}-q_{4}^{2}-q_{5}^{2}$$ invariant. Denoting the generators of this group by $J_{ij}$, we have, 
\begin{equation}
[J_{ij}, J_{kl}] = i(J_{ik}\eta_{jl} - J_{il}\eta_{jk} + J_{jl}\eta_{ik} - J_{jk}\eta_{il}) ~~~~~ i,j = 1, 2, 3, 4, 5 ~~,~~ \eta_{ij} = (+ + + - -) 
\label{eq3}
\end{equation}
\noindent The correspondance with the operators (\ref{eq2}) is given by, $$J_{ij} = \varepsilon_{ijk}L_{k}  ~~~~ for ~~i, j, k = 1, 2, 3$$ $$J_{i4} = K_{i}  ~~~~~~~~for ~~ i = 1,2,3~~~~~~~$$ $$J_{i5} = B_{i}  ~~~~~~~~for ~~ i = 1,2,3~~~~~~~$$  
\begin{equation}
J_{45} = H ~~~~~~~~~~~~~~~~~~~~~~~~~~~~~~~~~~~
\label{eq4}
\end{equation}

\noindent There are three spacial rotation operators $(L_{1}, L_{2}, L_{3})$, two boosts $(K_{i}, B_{i})$ corresponding to two times and a hamiltonian $(H)$. It may be observed that this $H$ in (\ref{eq2}) is half the energy of two harmonic oscillators, as also mentioned in \cite{pamd}. In section V we shall give an explanation for this half factor. The individual algebra satisfied by these operators is given by, $$[L_{i}, L_{j}] = i\varepsilon_{ijk}L_{k} ~~,~~ [L_{i}, H] = 0 ~~~~~~~~~~~~~~~~~~~~~~~~~~~~~~$$  $$[L_{i}, K_{j}] = i\varepsilon_{ijk}K_{k} ~,~~ [K_{i}, K_{j}] = -i\varepsilon_{ijk}L_{k}~~~~~~~~~~~~~~~~~~~$$  $$[L_{i}, B_{j}] = i\varepsilon_{ijk}B_{k} ~~,~~ [B_{i}, B_{j}] = -i\varepsilon_{ijk}L_{k}~~~~~~~~~~~~~~~~~~~~$$
\begin{equation}
[K_{i}, H] = iB_{i} ~~,~~ [B_{i}, H] = -iK_{i} ~~,~~ [K_{i}, B_{j}] = i\delta_{ij}H  ~~~
\label{eq5}
\end{equation}

\noindent The operators $L_{i}$ satisfy the $SU(2)$ algebra. Also, there are three sets of operators $(K_{1}, B_{1}, H)$, $(K_{2}, B_{2}, H)$ and $(K_{3}, B_{3}, H)$ that satisfy the hyperbolic $SU(1, 1)$ algebra. We will return to these issues later.

\section{Symplectic Transformations and Dirac's Representation}
\noindent Since the $3+2$ de-Sitter group is isomorphic to the four dimensional symplectic group $Sp(4)$, the algebra (\ref{eq5}) also serves to characterise the latter. This fact was utilised in \cite{kn, hkn} to illuminate the nature of the operators given in (\ref{eq2}). It was shown that the unitary transformations generated by these operators are translated into linear canonical transformations of the Wigner function for a pair of coupled oscillators. The corresponding group in this case is just $Sp(4)$. \\
\hspace*{2mm} In this section we provide an alternative approach that does not require the introduction of the Wigner function. The structure of the generators (\ref{eq2}) is deduced solely from properties of symplectic transformations. Starting from the matrix representation of the ten generators of $Sp(4)$, we construct the corresponding vector fields. From these fields, quadratic homogeneous polynomials involving four phase space variables are obtained. Finally, replacing the phase space variables by the oscillator variables, the ten generators in (\ref{eq2}) are reproduced. \\ 
\hspace*{2mm} A symplectic matrix $M$ is defined by the condition, 
\begin{equation}
MJM^{T} = J
\label{eq6}
\end{equation}

\noindent where $M^{T}$ is the transpose of $M$ and $J$ has the canonical form, 
\begin{equation}
J = 
\begin{pmatrix}
0 ~&~ I  \\
-I ~&~ 0 
\end{pmatrix}
\label{eq7}
\end{equation}

\noindent with $I$ being the identity matrix. For two pairs of canonical variables relevant for $Sp(4)$, $J$ takes the antisymmetric structure, 
\begin{equation}
J = 
\begin{pmatrix}
0 ~&~ 0 ~&~ 1 ~&~ 0  \\
0 ~&~ 0 ~&~ 0 ~&~ 1  \\
-1 ~&~ 0 ~&~ 0 ~&~ 0  \\
0 ~& -1 ~&~ 0 ~&~ 0  \\
\end{pmatrix}
\label{eq8}
\end{equation}
\noindent Introducing the generators $G_{i}$ of $Sp(4)$ we have, 
\begin{equation}
M = e^{-i\alpha_{i}G_{i}}
\label{eq9}
\end{equation}
\noindent where $G_{i}$ is a set of ten pure imaginary $4\times4$ matrices. The symplectic condition (\ref{eq6}) now implies that $G$ be either antisymmetric and commute with $J$ (first set) or, be symmetric and anticommute with $J$ (second set) \cite{kn}. \\  
\hspace*{2mm} Using the Pauli spin matrices and the $2\times2$ identity matrix it is possible to construct the relevant generators. The first set is given by, 
$$
L_{1} = \dfrac{i}{2}
\begin{pmatrix}
0 ~&~ \sigma_{1} \\
-\sigma_{1} ~&~ 0
\end{pmatrix}
~~,~~ L_{2} = \dfrac{1}{2}
\begin{pmatrix}
\sigma_{2} ~&~ 0\\
0 ~&~ \sigma_{2}
\end{pmatrix}
$$
\begin{equation}
L_{3} = \dfrac{i}{2}
\begin{pmatrix}
0 ~&~ \sigma_{3} \\
-\sigma_{3} ~&~ 0
\end{pmatrix}
~~,~~ H = \dfrac{i}{2}
\begin{pmatrix}
0 ~&~ I\\
-I ~&~ 0
\end{pmatrix}
\label{eq10}
\end{equation}
Likewise, the second set is given by, 
$$
K_{1} = \dfrac{i}{2}
\begin{pmatrix}
0 ~&~ \sigma_{3} \\
\sigma_{3} ~&~ 0
\end{pmatrix}
~~,~~ K_{2} = \dfrac{i}{2}
\begin{pmatrix}
I ~&~ 0\\
0 ~&~ -I
\end{pmatrix}
~~,~~K_{3} = -\dfrac{i}{2}
\begin{pmatrix}
0 ~&~ \sigma_{1} \\
\sigma_{1} ~&~ 0
\end{pmatrix}
$$
\begin{equation}
B_{1} = -\dfrac{1}{2}
\begin{pmatrix}
\sigma_{3}  ~&~ 0 \\
0 ~&~ -\sigma_{3} 
\end{pmatrix}
~~,~~ B_{2} = \dfrac{i}{2}
\begin{pmatrix}
0 ~&~ I\\
I ~&~ 0
\end{pmatrix}
~~,~~B_{3} = \dfrac{i}{2}
\begin{pmatrix}
\sigma_{1} ~&~ 0 \\
0 ~&~ -\sigma_{1}
\end{pmatrix}
\label{eq11}
\end{equation}
\noindent Using the result, 
\begin{equation}
[\sigma_{i}, \sigma_{j}] = 2i\varepsilon_{ijk}\sigma_{k}
\label{eq12}
\end{equation}
\noindent it is possible to show that the matrices (\ref{eq10}, \ref{eq11}) satisfy the algebra (\ref{eq5}). \\
\hspace*{2mm} We now show that the above matrices can be translated into oscillator variables that precisely reproduce the representation (\ref{eq2}). Let us consider the space $X = \rm I\!R^{2n} = \rm I\!R^{n} + \rm I\!R^{n}$  which has the symplectic form $w$,
\begin{equation}
w = \Sigma~dq_{i} \land dp_{i}
\label{eq13}
\end{equation}
\noindent Let $\xi$ be the vector field on $X$ and $i(\xi)$ denote the interior product by $\xi$. Then there is a smooth function $H$ determined up to a local constant such that \cite{gs}, 
\begin{equation}
i(\xi)w = dH
\label{eq14}
\end{equation} 
\noindent Hence, for $w$ defined in (\ref{eq13}), the vector field is given by, 
\begin{equation}
\xi_{H} = \Sigma ~ \bigg(\dfrac{\partial H}{\partial p_{i}}\dfrac{\partial}{\partial q_{i}} -\dfrac{\partial H}{\partial q_{i}}\dfrac{\partial}{\partial p_{i}}\bigg)
\label{eq15}
\end{equation}

\noindent The function $H$ is the hamiltonian and determines the evolution of the vector field (\ref{eq15}) from the differential equations, 
\begin{equation}
\dfrac{dq_{i}}{dt} = \dfrac{\partial H}{\partial p_{i}} ~~,~~ \dfrac{dp_{i}}{dt} = -\dfrac{\partial H}{\partial q_{i}}
\label{eq16}
\end{equation}
\noindent The poisson bracket of two functions $f_{1}$ and $f_{2}$ is defined as, 
\begin{equation}
\{f_{1}, f_{2}\} = -\xi_{f_{1}}f_{2}
\label{eq17}
\end{equation}
so that the familiar Hamilton's equations are obtained from (\ref{eq16}),
\begin{equation}
\dfrac{dq_{i}}{dt} = \{q_{i}, H\} ~~,~~ \dfrac{dp_{i}}{dt} = \{p_{i}, H\}
\label{eq18}
\end{equation}
Using the group property it can be shown that \cite{gs}, 
\begin{equation}
[\xi_{f_{1}}, \xi_{f_{2}}] = -\xi_{\{f_{1}, f_{2}\}}
\label{eq19}
\end{equation}
This is illustrated by taking a simple example. Consider the functions $f_{1}$ and $f_{2}$ to be the hamiltonians for the harmonic oscillator and free particle. respectively, 
\begin{equation}
f_{1} = \dfrac{1}{2}(p^{2} + q^{2}) ~~,~~ f_{2} = \dfrac{p^{2}}{2}
\label{eq20}
\end{equation}
Then the corresponding vector fields are obtained from (\ref{eq15}),
\begin{equation}
\xi_{f_{1}} = p\dfrac{\partial}{\partial q} - q\dfrac{\partial}{\partial p} ~~,~~\xi_{f_{2}} = p\dfrac{\partial}{\partial q}  
\label{eq21}
\end{equation}
The vector field corresponding to the Poisson bracket $\{f_{1}, f_{2}\}$ (defined in (\ref{eq17})) turns out to be, 
\begin{equation}
\xi_{\{f_{1}, f_{2}\}} = \xi_{qp} = q\dfrac{\partial}{\partial q} - p\dfrac{\partial}{\partial p}
\label{eq22}
\end{equation}
It is now easy to prove (\ref{eq19}) using (\ref{eq21}) and (\ref{eq22}). This shows an isomorphism between the algebra of quadratic homogeneous polynomials (in two variables) and that of their corresponding vector fields. The isomorphism can be pushed further to include the symplectic algebra. To do this we first write the appropriate matrices.  \\
 \hspace*{2mm} For the function $f_{1}$ the equations of motion (\ref{eq16}) are given by, 
\begin{equation}
\dfrac{dq}{dt} = \dfrac{\partial f_{1}}{\partial p} = p ~~,~~ \dfrac{dp}{dt} = -\dfrac{\partial f_{1}}{\partial q} = -q
\label{eq23}
\end{equation} 
 In matrix notation, 
 \begin{equation}
\dfrac{d}{dt} 
\begin{pmatrix}
q \\
p
\end{pmatrix}
= 
\begin{pmatrix}
0 ~&~ 1 \\
-1 ~&~ 0
\end{pmatrix}
\begin{pmatrix}
q \\
p
\end{pmatrix}
\label{eq24} 
 \end{equation}
 Thus the function $f_{1}$ corresponds to the vector field $\xi_{f_{1}}$ and the matrix $\begin{pmatrix}
 0 ~&~ 1 \\
 -1 ~&~ 0
 \end{pmatrix}
$ belonging to the Lie algebra of the symplectic group $Sp(2)$. Likewise we can find the matrices corresponding to $f_{2}$ and $\{f_{1}, f_{2}\}$. Collecting all results, 
\begin{equation}
M_{f_{1}} = 
\begin{pmatrix}
0 ~&~ 1 \\
-1 ~&~ 0
\end{pmatrix}
~~,~~ 
M_{f_{2}} = 
\begin{pmatrix}
0 ~&~ 1 \\
0 ~&~ 0
\end{pmatrix}
~~,~~ 
M_{\{f_{1}, f_{2}\}} = 
\begin{pmatrix}
1 ~&~ 0 \\
0 ~&~ -1
\end{pmatrix}
\label{eq25}
\end{equation}
It is easy to see that the algebra closes,
\begin{equation}
 [M_{f_{1}}, M_{f_{2}}] = M_{\{f_{1}, f_{2}\}}
\label{eq26}
\end{equation}
Associating a self adjoint operator $f_{q}$ to each homogeneous quadratic polynomial $f$, it is possible to construct the map $f \longrightarrow -if_{q}$ such that Poisson brackets get replaced by commutators. We then have an isomorphim between the commutator algebra of the quadratic polynomials and the Lie algebra of the symplectic group. Finally, the quadratic polynomials may be equivalently represented in terms of the oscillator variables. Following this approach Dirac's representation (\ref{eq2}) is derived starting from the generators (\ref{eq10}, \ref{eq11}) of the symplectic group $Sp(4)$. \\
\hspace*{2mm} Let us explicitly work out one specific example. Take $L_{1}$ from the set (\ref{eq10}).
\begin{equation}
L_{1} = \dfrac{i}{2}
\begin{pmatrix}
0 ~&~ \sigma_{1} \\
-\sigma_{1} ~&~ 0
\end{pmatrix}
= \dfrac{i}{2}
\begin{pmatrix}
0 ~&~ 0 ~&~ 0 ~&~ 1 \\
0 ~&~ 0 ~&~ 1 ~&~ 0 \\
0 ~& -1 ~&~ 0 ~&~ 0 \\
-1 ~&~ 0 ~&~ 0 ~&~ 0 
\end{pmatrix}
\label{eq27}
\end{equation}
The corresponding vector field is given by, 
\begin{equation}
\xi_{L_{1}} = \dfrac{i}{2}\bigg(p_{2}\dfrac{\partial}{\partial q_{1}} + p_{1}\dfrac{\partial}{\partial q_{2}} - q_{2}\dfrac{\partial}{\partial p_{1}} - q_{1}\dfrac{\partial}{\partial p_{2}}\bigg)
\label{eq28}
\end{equation} 
 The homogeneous quadratic polynomial leading to (\ref{eq28}) is obtained from (\ref{eq15}), 
 \begin{equation}
 L_{1} = \dfrac{i}{2}(p_{1}p_{2} + q_{1}q_{2})
 \label{eq29}
 \end{equation}
 As a cross check we reproduce (\ref{eq27}) from (\ref{eq29}). The equations of motion (\ref{eq16}) are,
 \begin{equation}
 \dfrac{dq_{1}}{dt} = \dfrac{\partial L_{1}}{\partial p_{1}} = \dfrac{i}{2}p_{2} ~,~ \dfrac{dq_{2}}{dt} = \dfrac{i}{2}p_{1} ~,~ \dfrac{dp_{1}}{dt} = -\dfrac{i}{2}q_{2} ~,~ \dfrac{dp_{2}}{dt} = -\dfrac{i}{2}q_{1} 
 \label{eq30}
 \end{equation}
 so that, 
 \begin{equation}
 \dfrac{d}{dt}  
 \begin{pmatrix}
 q_{1} \\
 q_{2} \\
 p_{1}  \\
 p_{2}
 \end{pmatrix}
 = \dfrac{i}{2}
 \begin{pmatrix}
0 ~&~ 0 ~&~ 0 ~&~ 1 \\
0 ~&~ 0 ~&~ 1 ~&~ 0 \\
0 ~& -1 ~&~ 0 ~&~ 0 \\
-1 ~&~ 0 ~&~ 0 ~&~ 0 
 \end{pmatrix}
 \begin{pmatrix}
 q_{1} \\
 q_{2} \\
 p_{1}  \\
 p_{2}
 \end{pmatrix}
 \label{eq31}
 \end{equation}
thereby yielding the matrix (\ref{eq27}). Next, replacing $L_{1}$ by $-iL_{1}$ and substituting $q, p$ by corresponding operators, we obtain the operator version of (\ref{eq29}), 
\begin{equation}
\hat{L_{1}} = \dfrac{1}{2}(\hat{p_{1}}\hat{p_{2}} + \hat{q_{1}}\hat{q_{2}})
\label{eq32}
\end{equation} 
which is already hermitian since, 
\begin{equation}
[\hat{q_{i}}, \hat{p_{j}}] = i\delta_{ij} ~~,~~ [\hat{q_{i}}, \hat{q_{j}}] = [\hat{p_{i}}, \hat{p_{j}}] = 0
\label{eq33}
\end{equation} 
Introducing the ladder (creation and destruction) operators for oscillators, 
\begin{equation}
\hat{q_{i}} = \dfrac{1}{\sqrt{2}}(a_{i} + a^{\dagger}_{i}) ~~,~~ \hat{p_{i}} = -\dfrac{i}{\sqrt{2}}(a_{i} - a^{\dagger}_{i})
\label{eq34}
\end{equation} 
 we finally obtain from (\ref{eq32}), 
 \begin{equation}
 \hat{L_{1}} = \dfrac{1}{2}(a^{\dagger}_{1}a_{2} + a^{\dagger}_{2}a_{1})
 \label{eq35}
 \end{equation}
 which reproduces the first relation in (\ref{eq2}). \\
 \hspace*{2mm} In cases, unlike (\ref{eq32}), which are not hermitian, one has to just add the hermitian conjugate. As an example, consider $K_{2}$ from (\ref{eq11}),
 \begin{equation}
 K_{2} = \dfrac{i}{2}
 \begin{pmatrix}
1 ~&~ 0 ~&~ 0 ~&~ 0 \\
0 ~&~ 1 ~&~ 0 ~&~ 0 \\
0 ~&~ 0 ~& -1 ~&~ 0 \\
0 ~&~ 0 ~&~ 0 ~& -1 
\end{pmatrix}
 \label{eq36}
 \end{equation}
 The vector field is, 
 \begin{equation}
\xi_{K_{2}} = \dfrac{i}{2}\bigg(q_{1}\dfrac{\partial}{\partial q_{1}} + q_{2}\dfrac{\partial}{\partial q_{2}} - p_{1}\dfrac{\partial}{\partial p_{1}} - p_{1}\dfrac{\partial}{\partial p_{2}}\bigg)
\label{eq37}
\end{equation} 
 and the corresponding polynomial follows from (\ref{eq15}),
 \begin{equation}
 K_{2} = \dfrac{i}{2}(q_{1}p_{1} + q_{2}p_{2})
 \label{eq38}
 \end{equation}
 The operator version is given by, 
 \begin{equation}
 \hat{K_{2}} = \dfrac{1}{4}(\hat{q_{1}}\hat{p_{1}} + \hat{p_{1}}\hat{q_{1}} + \hat{q_{2}}\hat{p_{2}} + \hat{p_{2}}\hat{q_{2}})
 \label{eq39}
 \end{equation}
 where we have included the hermitian conjugate so that $\hat{K_{2}}$ is hermitian. Using the basic brackets (\ref{eq33}), this reduces to, 
 \begin{equation}
 \hat{K_{2}} = \dfrac{1}{2}(\hat{q_{1}}\hat{p_{1}} + \hat{p_{2}}\hat{q_{2}})
 \label{eq40}
 \end{equation}
 Translating into the oscillator variables (\ref{eq34}), this reproduces the corresponding result in (\ref{eq2}). \\ 
 \hspace*{2mm} In this manner it is possible to derive the complete representation (\ref{eq2}) starting from (\ref{eq10}, \ref{eq11}). 
 \section{Generalised Jordan-type map}
\noindent It is useful to note that the matrices (\ref{eq10}, \ref{eq11}) which eventually led to the representation (\ref{eq2}) are all composed of the Pauli matrices and the identity matrix. This suggests the possibility of obtaining (\ref{eq2}) directly from the Pauli and identity matrices. Here we show how this can be done. \\ 
\hspace*{2mm} Before coming to the actual solution, let us first show that there are other possible representations of the de-Sitter group in terms of oscillator variables, which are different from (\ref{eq2}). This is based on Jordan's map,
\begin{equation}
\hat{M_{i}} = \dfrac{1}{2}a^{\dagger}M_{i}a
\label{neweq41}
\end{equation}
which states that the algebra of the operators $\hat{M_{i}}$ is isomorphic to that of the matrices $M_{i}$, where $a$ and $a^{\dagger}$ are the usual ladder operators. In other words if the $M_{i}$ satisfy, 
\begin{equation}
[M_{i}, M_{j}] = \omega_{ijk}M_{k}
\label{neweq42}
\end{equation} 
where $\omega_{ijk}$ are some structure constants, then, 
\begin{equation}
[\hat{M_{i}}, \hat{M_{j}}] = \omega_{ijk}\hat{M_{k}}
\label{neweq43}
\end{equation}
so that $\hat{M_{i}}$ may be interpreted as an operator representation of the matrices $M_{i}$. Since we have earlier provided the ten generators of $Sp(4)$ in matrix form (\ref{eq10}, \ref{eq11}), it is straightforward to compute their corresponding operator versions from (\ref{neweq41}). Thus for $L_{1}$ (\ref{eq10}), we have, 
\begin{equation}
\hat{L_{1}} = \dfrac{i}{4}(a_{1}^{\dagger}a_{4}+a_{2}^{\dagger}a_{3}-a_{3}^{\dagger}a_{2}-a_{4}^{\dagger}a_{1})
\label{neweq44}
\end{equation} 
Likewise, it is possible to obtain the relevant operators for all the matrices (\ref{eq10}, \ref{eq11}). Clearly, these will satisfy the de-Sitter algebra (\ref{eq5}).\\
\hspace*{2mm} The structure found in this way is obviously different from the representation (\ref{eq2}). It now involves four coupled oscillators instead of two. Among other things, the representation (\ref{eq2}) is minimal in the sense that, to construct ten independent quadratic operators in the ladder variables, the minimum number of the oscillators needed is two. It is now clear that to reproduce (\ref{eq2}) in this fashion, we have  to work with ($2\times2$) matrices instead of ($4\times4$)ones. As indicated, the obvious choice would be the Pauli matrices and the ($2\times2$) identity matrix.\\
\hspace*{2mm} There is a well known representation of the $SU(2)$ generators in terms of the oscillator (ladder) operators. This is called the Jordan-Schwinger map. It is a special version of (\ref{neweq41}) with $M_{i}$ replaced by the Pauli matrices and $\hat{M_{i}}$ by $L_{i}$, the angular momentum operator,
\begin{equation}
L_{i} = \dfrac{1}{2}~a^{\dagger}\sigma_{i}a
\label{eq41}
\end{equation}  
from which follows, 
\begin{equation}
[L_{i}, L_{j}] = i\varepsilon_{ijk}L_{k}
\label{eq42}
\end{equation}
 which is the first relation in (\ref{eq5}). Indeed, using the explicit structure of the Pauli matrices, $L_{1}, L_{2}$ and $L_{3}$ exactly reproduce the corresponding relations in (\ref{eq2}). \\ 
 \hspace*{2mm} The point is that (\ref{eq41}) is not the only possible way of expressing a hermitian operator in terms of a product of  the Pauli matrices and quadratic functions of the oscillator variables. For instance, we can have the form, 
 \begin{equation}
 W_{i} = \dfrac{1}{4}(a^{\dagger}\sigma_{i}a^{\dagger} + a\sigma_{i}a)
 \label{eq43}
 \end{equation}
 which, like (\ref{eq41}), is also hermitian. Inserting the explicit form for the $\sigma$-matrices, we find, 
 \begin{equation}
 W_{1} = K_{3} ~~,~~ W_{3} = -K_{1}
 \label{eq44}
 \end{equation}
Note that $W_{2} = 0$ due to reasons of symmetry. Another hermitian construction, similar to (\ref{eq43}), is given by, 
\begin{equation}
Z_{i} = \dfrac{i}{4}(a^{\dagger}\sigma_{i}a^{\dagger} - a\sigma_{i}a)
\label{eq45}
\end{equation}  
 We find, 
 \begin{equation}
 Z_{1} = B_{3} ~~,~~ Z_{3} = -B_{1} 
 \label{eq46}
 \end{equation}
while $Z_{2} = 0$. This exhausts possible (quadratic) hermitian combinations involving the Pauli matrices. Now we consider the identity matrix. \\
\hspace*{2mm} The analogues of (\ref{eq43}) and (\ref{eq45}) involving the identity matrix are given by,
\begin{equation}
W = \dfrac{1}{4}(a^{\dagger}_{i}a^{\dagger}_{i} + a_{i}a_{i})
\label{eq47}
\end{equation}  
 and, 
\begin{equation}
Z = \dfrac{i}{4}(a^{\dagger}_{i}a^{\dagger}_{i} - a_{i}a_{i})
\label{eq48}
\end{equation} 
 which yield, 
 \begin{equation}
 W = -B_{2} ~~,~~ Z = K_{2}
 \label{eq49}
 \end{equation}
 Finally, we have the structure,
 \begin{equation}
 H = \dfrac{1}{4}(a^{\dagger}_{i}a_{i} + a_{i}a^{\dagger}_{i})
 \label{eq50}
 \end{equation}
 which reproduces the corresponding expression in (\ref{eq2}). Thus the complete representation (\ref{eq2}) may be deduced by using a generalisation of the Jordan map to include all possible quadratic combinations of the oscillator operators sandwitching the Pauli and identity matrices. These combinations have to be hermitian.
 \section{Physical Interpretation}
 In this section we discuss certain features of the representation (\ref{eq2}) that manifest a physical structure. In the cyclic sector comprising $L_{1}, L_{2}, L_{3}$ and $H$, while $L_{1}$ and $L_{2}$ have coupled expressions, $L_{3}$ and $H$ correspond to decoupled oscillators. Indeed it is possible to reproduce these expressions by considering a two dimensional oscillator as a pair of chiral oscillators, one rotating in the clockwise direction while the other in the anticlockwise direction. \\ 
 \hspace*{2mm} The lagrangian for a one-dimensional harmonic oscillator is given by, 
 \begin{equation}
 L = \dfrac{1}{2}(\dot{q}^{2} - q^{2})
 \label{eq51}
\end{equation}  
 By introducing an additional variable it can be converted into its first order form \cite{bg}, 
 \begin{equation}
 L_{\pm} = \dfrac{1}{2}(\pm ~\varepsilon_{\alpha\beta}q_{\alpha}\dot{q_{\beta}} - q^{2}_{\alpha})
 \label{eq52}
 \end{equation}
 where $\alpha = 1, 2$ is an internal index with $\epsilon_{12} = 1$. Eliminating either $q_{1}$ or $q_{2}$ in favour of the other yields (\ref{eq51}) with $q = q_{1}$ or $q_{2}$. The lagrangians $L_{\pm}$ are a pair of chiral oscillators with the sign of the first term determining their chirality. \\
 \hspace*{2mm} Although the lagrangians (\ref{eq52}) each seem to have two degrees of freedom, only one of them is independent due to the presence of constraints which imply the symplectic structure,
 \begin{equation}
 \{q_{\alpha}, q_{\beta}\} = \mp~ \varepsilon_{\alpha\beta}
 \label{eq53}
 \end{equation}
 for $L_{\pm}$. The corresponding hamiltonians are identical due to the first order nature of (\ref{eq52}), 
 \begin{equation}
 H_{\pm} = \dfrac{1}{2}(q_{1}^{2} + q_{2}^{2}) = \tilde{H}
 \label{eq54}
 \end{equation}
 The angular momentum is given by, 
 \begin{equation}
 J_{\pm} = \varepsilon_{\alpha\beta}q_{\alpha}p_{\beta}
 \label{eq55}
 \end{equation}
 where the canonical momenta $p_{\beta}$ is, 
 \begin{equation}
 p_{\beta} = \dfrac{\partial L_{\pm}}{\partial \dot{q_{\beta}}} = \pm ~ \dfrac{1}{2} \varepsilon_{\rho\beta}q_{\rho}
 \label{eq56}
 \end{equation}
 Thus, 
 \begin{equation}
 J_{\pm} = \pm~ \dfrac{1}{2}q_{\alpha}^{2} = \pm~ \tilde{H}
 \label{eq57}
 \end{equation}
 so that the angular momenta have the same magnitude, but differ in sign, a consequence of chirality. \\
 \hspace*{2mm} Let us now show explicitly how $L_{+}$ and $L_{-}$ may be combined to yield a two dimensional oscillator. Consider the following combination of $L_{+}(q_{\alpha})$ and $L_{-}(r_{\alpha})$,
 \begin{equation}
 L = L_{+}(q_{\alpha}) + L_{-}(r_{\alpha}) = \dfrac{1}{2}\varepsilon_{\alpha\beta}q_{\alpha}\dot{q_{\beta}} - \dfrac{1}{2}q_{\alpha}^{2} - \dfrac{1}{2}\varepsilon_{\alpha\beta}r_{\alpha}\dot{r_{\beta}} - \dfrac{1}{2}r_{\alpha}^{2}
 \label{eq58}
 \end{equation}
 Introducing new variables, 
 \begin{equation}
 q_{\alpha} + r_{\alpha} = Q_{\alpha} ~~,~~ q_{\alpha} - r_{\alpha} = R_{\alpha}
 \label{eq59}
 \end{equation}
 we get,
\begin{equation}
 L = \dfrac{1}{2}\varepsilon_{\alpha\beta}R_{\alpha}\dot{Q_{\beta}} - \dfrac{1}{4}(Q_{\alpha}^{2} + R_{\alpha}^{2})
\label{eq60}
\end{equation} 
where a total time derivative term has been dropped. Eliminating either $R$ or $Q$ in favour of the other yields, 
\begin{equation}
L = \dfrac{1}{4}(\dot{Z_{\alpha}}^{2} - Z_{\alpha}^{2})  ~~;~~ Z_{\alpha} = R_{\alpha} ~~or~~ Q_{\alpha}
\label{eq61}
\end{equation} 
 which is the standard lagrangian for the two dimensional oscillator but with an averall normalisation of $\dfrac{1}{2}$. \\ 
 \hspace*{2mm} From (\ref{eq53}, \ref{eq54}) we find that the hamiltonians for the chiral oscillators have identical expressions as normal oscillators.\hyperlink{footnote}{$^{1}$} Hence, accounting for the $\dfrac{1}{2}$ factor mentioned above, the final hamiltonian has the form \\
 \begin{minipage}[b]{1\textwidth}
 ~~~~~~~~~~~~~~~~~~\\
 \rule{15mm}{0.005pt}\\
\hypertarget{footnote}{$^{1}$}\footnotesize{Note that the canonical pairs ($q, p$) for $L_{+}$ and $L_{-}$ are ($q_{2}, q_{1}$) and ($q_{1}, q_{2}$), respectively.}
\end{minipage} 

 \begin{equation}
 H = \dfrac{1}{2}(H_{+} + H_{-}) = \dfrac{1}{2}\bigg((a^{\dagger}_{1}a_{1} + \dfrac{1}{2}) + (a^{\dagger}_{2}a_{2} + \dfrac{1}{2})\bigg)   = \dfrac{1}{2}(a^{\dagger}_{1}a_{1} + a^{\dagger}_{2}a_{2} + 1)
 \label{eq62}
 \end{equation}
 while the angular momentum, which is the difference between $H_{+}$ and $H_{-}$, becomes, 
 \begin{equation}
 J = \dfrac{1}{2}\bigg((a^{\dagger}_{1}a_{1} + \dfrac{1}{2}) - (a^{\dagger}_{2}a_{2} + \dfrac{1}{2})\bigg) = \dfrac{1}{2}(a^{\dagger}_{1}a_{1} - a^{\dagger}_{2}a_{2})
 \label{eq63}
 \end{equation}
These expressions match with the corresponding ones in (\ref{eq2}) (i.e. $H$ and $L_{z}$). Thus two chiral oscillators rotating in th $x-y$ plane in opposite (clockwise and anticlockwise) directions effectively yield the hamiltonian ($H$) and angular momentum ($L_{z}$ or $L_{3}$) given in (\ref{eq2}). From (\ref{eq62}) it is seen that all eigenvalues of the hamiltonian are positive with the minimum being $\dfrac{1}{2}$, corresponding to the zero point energy of the two chiral oscillators. If they had been usual oscillators then this minimum would have been 1. Likeiwse, the $z$-component of the angular momentum (\ref{eq63}) can have both integral and half integral values with the sign determined by the difference in the number of clockwise and anticlockwise rotating chiral oscillators. \\
\hspace*{2mm} A similar analysis may be done to understand the physical content of the remaining generators containing only uncoupled terms (like $K_{1}, K_{2}, B_{1}, B_{2}$). Incidentaly, ($K_{1}, B_{1}, H$) and ($K_{2}, B_{2}, H$) satisfy an algebra isomorphic to $SU(1,1)$, as easily checked from (\ref{eq5}). Thus the starting point is to consider the basic matrices defining this algebra.\\ 
\hspace*{2mm}Consider the following $2\times2$ matrices,
\begin{equation}
S_{1} = \dfrac{i}{2}
\begin{pmatrix}
1 ~&~ 0 \\
0 ~& -1
\end{pmatrix}
~~,~~ 
S_{2} = \dfrac{i}{2}
\begin{pmatrix}
0 ~&~ 1 \\
-1 ~&~ 0
\end{pmatrix}
~~,~~ 
S_{3} = \dfrac{i}{2}
\begin{pmatrix}
0 ~&~ 1 \\
1 ~&~ 0
\end{pmatrix}
\label{eq64}
\end{equation}
 which satisfy the $SU(1,1)$ algebra,
 \begin{equation}
 [S_{1}, S_{2}] = iS_{3} ~~,~~ [S_{2}, S_{3}] = iS_{1} ~~,~~ [S_{3}, S_{1}] = -iS_{2}
 \label{eq65}
 \end{equation}
Following the technique discussed in sction III, the vector field corresponding to $S_{1}$ is given by,
\begin{equation}
\xi_{S_{1}} = \dfrac{i}{2}\bigg(q\dfrac{\partial}{\partial q} - p\dfrac{\partial}{\partial p}\bigg)
\label{eq66}
\end{equation} 
Using (\ref{eq15}), the relevant polynomial generating this vector is found to be, 
\begin{equation}
S_{1} = \dfrac{i}{2}(qp)
\label{eq67}
\end{equation} 
and the corresponding hermitian operator is,  
 \begin{equation}
\hat{S_{1}} = -\dfrac{1}{4}(\hat{q}\hat{p} + \hat{p}\hat{q})
\label{eq68}
\end{equation}
 Translating into oscillator variables, 
  \begin{equation}
\hat{S_{1}} = \dfrac{i}{4}(a^{\dagger2} - a^{2})
\label{eq69}
\end{equation}
 Similarly $S_{2}$ and $S_{3}$ translate into,
 \begin{equation}
\hat{S_{2}} = \dfrac{1}{4}(a^{\dagger}a + aa^{\dagger}) ~~,~~ \hat{S_{3}} = -\dfrac{1}{4}(a^{\dagger2} + a^{2})
\label{eq70}
\end{equation}
 The operators $\hat{S_{1}},\hat{S_{2}}$ and $\hat{S_{3}}$ satisfy the same algebra as (\ref{eq65}). \\ 
 \hspace*{2mm} We may now identify, with appropriate modifications, the set $\hat{S_{1}}, \hat{S_{2}}, \hat{S_{3}}$ with $K_{2}, H$ and $B_{2}$, respectively, as defined in (\ref{eq2}). One has to just consider two independent oscillators instead of a single one. Then the transition of $\hat{S_{1}}$ and $\hat{S_{3}}$ to $K_{2}$ and $B_{2}$ is straightforward. For $\hat{S_{2}}$, taking two independent oscillators yields, 
 $$ \hat{S_{2}}|_{1} +  \hat{S_{2}}|_{2} = \dfrac{1}{4}(a^{\dagger}_{1}a_{1} + a_{1}a^{\dagger}_{1} + a^{\dagger}_{2}a_{2} + a_{2}a^{\dagger}_{2} )$$
 \begin{equation}
 = \dfrac{1}{2}(a^{\dagger}_{1}a_{1} + a_{2}a^{\dagger}_{2})~~~~
 \label{eq71}
 \end{equation}
 obtained on using basic commutation relations. The final form is just $H$ in (\ref{eq2}). Incidently, the distinctive roles of $SU(2)$ and $SU(1)$ groups in interferometry has been discussed in \cite{ymk}. Here we see another example in the case of the de-Sitter representation (\ref{eq2}).
 \section{Conclusion}
\noindent In this paper we have looked at the remarkable representation (\ref{eq2}) of the $3+2$ de-Sitter group found by Dirac \cite{pamd}. Although this work was done in 1963, its importance can be gauged from the fact that it is still relevant in contemporary physics. Among other applications \cite{kn}, this construction became the basic mathematical language for discussing two mode squeezed states in quantum optics \cite{hkn,hkny,ymk}. \\
\hspace*{2mm}Since the algebra of the $3+2$ de-Sitter group is isomorphic to the four dimensional symplectic group $Sp(4)$, it is usually customary to analyse the representation (\ref{eq2}) through this group. This is because the symplectic group, being the group of canonical transformations, provides a natural connection between physics and mathematics.\\
\hspace*{2mm} The matrices representing the generators of $Sp(4)$ were translated into homogeneous quadratic polynomials involving a pair of phase space variables ($q_{1}, p_{1}, q_{2}, p_{2}$). The step by step sequence of this construction was discussed. Once the polynomial was obtained, it was straightforward to express these in terms of the oscillator (ladder) operators using the standard expressions. In this way the complete representation (\ref{eq2}) was obtained.\\
\hspace*{2mm} It may be recalled that, in refs.\cite{kn,hkn,hkny}, the set (\ref{eq2}) was derived by using the ten matrices of $Sp(4)$ and performing canonical transformations on Wigner's phase space function involving a pair of canonical variables. Our derivation is more geometrical and does not need the Wigner function. While the connection of this function with density matrices is well known [~], the present analysis reveals its geometrical origin. Indeed the vector fields found here by symplectic considerations agree with those found by Wigner's phase space approach \cite{kn}.\\
 \hspace*{2mm}Motivated by the use of Pauli and identity matrices in the building of the $Sp(4)$ representation, we have given a simple derivation of (\ref{eq2}) that generalises the well known Jordan-Schwinger map for $SU(2)$ matrices. By constructing the most general quadratic hermitian combinations of the ladder operators with the Pauli and identity matrices, this result was obtained. Effectively, the set (\ref{eq2}) may be interpreted as this generalised Jordan type map for the de-Sitter group. Since $SU(2)$ is a subgroup of the de-Sitter group, the usual Jordan map is just a subset of this map that is restricted to the three operators $L_{1}, L_{2}$ and $L_{3}$ in (\ref{eq2}).\\
 \hspace*{2mm}A physical interpretation of the results has also been provided. It is easy to see from (\ref{eq2}) that there are some operators that involve a coupling of the pair of oscillators, but some are uncoupled. The uncoupled set $L_{3}$ and $H$ was shown to fit in the interpretation of regarding ($a_{1}, a_{1}^{\dagger}$) and ($a_{2}, a_{2}^{\dagger}$) as the ladder operators corresponding to chiral oscillators rotating in clockwise and anticlockwise directions. Furthermore, this interpretation explains the appearance of the half factor in the hamiltonian $H$ in (\ref{eq2}). Also, some other operators of (\ref{eq2}) were shown to be a simple addition of the corresponding operators for two independent oscillators whose ladder operators are used to define the $SU(1,1)$ representation. \\
 \hspace*{2mm} The techniques developed here may be generalised in a systematic way to higher dimensions. For example, knowing the twenty one generators of $Sp(6)$ ($Sp(2n)$ has $2n^{2}+n$ generators) it would be possible to build a similar representation as (\ref{eq2}) comprising three coupled oscillators. Like (\ref{eq2}), this would also be the optimal or minimal representation since the number of independent quadratic functions of $a_{i}, a_{i}^{\dagger}~(i = 1, 2, 3)$ is exactly twentyone. Similarly, with more operators at hand, other generalised Jordan type maps analogous to (\ref{eq43}) and (\ref{eq45}) can be constructed. It is known that the Wigner phase space approach gets connected to other formalisms (like the Schoedinger approach) through density matrices. Our symplectic analysis provides another route, which is complimentary to the convensional group theoretic approach \cite{kn}. This aspect may be studied further by looking at specific problems in quantum mechanics.

 ~~~~~~~~~~~~~~~~~~~~~~~~~~~~~~~~~~~~~~ \\
 ~~~~~~~~~~~~~~~~~~~~~~~~~~~~~~~\\
\noindent {\bf{Acknowledgements}}: I thank Ankur Srivastav for typing the manuscript. 


\begin{thebibliography}{99}


\bibitem{pamd}P.A.M. Dirac, A Remarkable Representation of the 3 + 2 de Sitter Group \href{https://doi.org/10.1063/1.1704016}{J. Math. Phys. \underline{4}, 901 (1963).} \\

\bibitem{hpr}H.P. Robertson, \href{https://doi.org/10.1103/RevModPhys.5.62}{Rev. Mod. Phys. \underline{5}, 62(1933).}

\bibitem{epw1}E.P. Wigner, \href{ https://doi.org/10.1073/pnas.36.3.184}{Proc. Nat. Acad. Sci. U.S.A. \underline{36}, 184(1950).}, \href{https://www.jstor.org/stable/1968551}{Ann. Math. \underline{40}, 149(1939).}

 

\bibitem{kn}For a good collection of applications, see, Y.S. Kim amd M.E. Noz, Phase Space Picture of Quantum Mechanics (Group Theoretical Approach) \href{https://doi.org/10.1142/1197 }{Lecture Notes in Physics Series-Vol. 40, World Scientific, 1991.}

\bibitem{epw2}E.P. Wigner, \href{https://doi.org/10.1103/PhysRev.40.749}{Phys. Rev. \underline{40}, 749(1932).}

\bibitem{hkn}D. Han, Y.S. Kim, M.E. Noz, Phys. Rev. \underline{A41}, 6233(1990) \href{}{}

\bibitem{hkny}D. Han, Y.S. Kim, M.E. Noz and L.Yeh, Jour. Math. Phys. \underline{34}, 5493(1993) \href{}{}


\bibitem{ymk}B. Yurke, S.L. McCall and J.R. Klauder,  \href{https://doi.org/10.1103/PhysRevA.33.4033}{Phys. Rev. \underline{A 33}, 4033(1986). }

\bibitem{gs}V. Guillemin and S. Sternberg, \textit{Symplectic Techniques in physics},  \href{http://www-spires.fnal.gov/spires/find/books/www?cl=QC20.7.D52G945::1984}{Cambridge University Press, 1984.}

\bibitem{bg}R. Banerjee and S. Ghosh,  \href{https://doi.org/10.1088/0305-4470/31/36/002}{ J. Phys. \underline{A33}, 4033(1986).}


\end{thebibliography}
\end{document}